# Neutron scattering patterns show Superconductivity in FeTe$_{0.5}$Se$_{0.5}$ likely results from itinerant electron fluctuations.

H. A. Mook[1], M.D. Lumsden[2], A.D. Christianson[2], Brian C. Sales[3], Rongying Jin[3], Michael A. McGuire[3], Athena Sefat[3], D. Mandrus[3], S.E. Nagler[2], T. Egami [3,4] and Clarina de la Cruz [2,4]

1. Neutron Sciences Directorate, Oak Ridge National Laboratory, Oak Ridge, Tennessee 37831, USA.

2. Neutron Sciences Division, Oak Ridge National Laboratory, Oak Ridge, Tennessee 37831, USA.

3. Materials Science and Technology Division, Oak Ridge National Laboratory, Oak Ridge, Tennessee 37831, USA

4. Department of Physics and Astronomy, The University of Tennessee, Knoxville, Tennessee 37996-1200, USA

**The discovery of the Fe pnictide superconductors generated great interest as the structure consists of planes of a magnetic material quite similar to the cuprate superconductors[1,2,3]. Fe(Te$_{0.5}$Se$_{0.5}$) is a particularly simple system whose planes are isostructural to the FeAs layers found in the originally discovered superconductors of this type[4,5,6,7,8]. We report here neutron scattering measurements on this material**

**that provide an understanding of the superconductivity. Since the information about the Fermi surface is available both from photoemission and band structure calculations for FeTe, FeSe and other related materials[9,10,11,12,13] the neutron spectra can be used to see if the itinerant electron picture is valid. The results are consistent with a picture where there are both electron and hole Fermi surfaces that make exact $(\Pi, \Pi)$ transitions possible. This would normally favor either a spin or charge density wave state[10,13]. However, our measurements show the extent of the region where $(\Pi, \Pi)$ transitions take place and demonstrate that there are a much larger number of transitions near $(\Pi, \Pi)$. The near $(\Pi, \Pi)$ transitions are observed both above and below Tc and are expected to be strongly pairing[10]. The superconductivity can be attributed to these excitations while the exactly $(\Pi, \Pi)$ transitions produce the narrow resonance[14,15,16] excitation that appears below Tc.**

The recently discovered ferrous superconductors were found when F was doped into LaFeAsO.[1] If La is replaced by magnetic lanthanides (e.g. Ce, Pr, Nd, Sm), $T_c$ increases as high as 55 K[3]. Remarkably the lanthanide oxide/fluoride charge reservoir layer found in the original superconductors turns out not to be necessary for superconductivity, and is absent in the Fe(TeSe) systems[4,5,6]. Density functional calculations have been made for the Fe(TeSe) system and calculations of the electron-phonon coupling show that FeSe is not an electron-phonon superconductor, but within a spin-fluctuation driven picture FeTe with doping would give the highest Tc of the Fe(TeSe) system[7,8]. The calculations give two intersecting elliptical cylindrical electron Fermi surfaces at the zone corner which are compensated by lower velocity hole sections at the zone center. Transitions between these Fermi surface sections that have opposite sign order parameters would result from the s± pairing[10] state. These transitions would appear around the ($\Pi,\Pi$) reciprocal position using tetragonal lattice notation.. However, since the Fermi surfaces of the doped material are unlikely to be cylinders of equal radius, transitions other than those at exactly ($\Pi,\Pi$) are expected. These are strongly superconducting and if the neutron peak is wide these have a much larger weight so that superconductivity is favored[10]. Recent photoemission measurements show a Fermi surface different than the calculations in that the surface shows a more complex shape[13]. However, this would produce broad spectra in agreement with the neutron measurements.

The structure of the system is shown in Fig.1a. This is the α phase, which is the phase of interest. This phase ($Fe_{1+y}Te_xSe_{1-x}$) exists for values of x between 0 and 1. The Fe(1) sites of the α phase structure are fully occupied, and excess Fe (y) partially occupy the interstitial sites marked as Fe(2). For x= 1, the minimum value of y is about 0.06, but for smaller values of x (increasing Se content), the value of y approaches 0 [7,17,18]. Large crystals can be grown via a Bridgman method for 0.5 <x< 1.[18] Although traces of superconductivity with $T_c \approx$ 14 K are found in all of the crystals with x < 0.9, bulk superconductivity is only found in crystals with an average composition of $FeTe_{0.5}Se_{0.5}$[18].

Heat capacity, resistivity and magnetic susceptibility give a $T_c$ of about 14K (Fig 1b and Fig 1c). Powder x-ray diffraction and EDX analysis of slices cut from various regions of the crystal indicates the crystal consists only of the α phase but with separate distinct regions. These have Te/Se ratios of approximately 55/45 and 35/65 with at least 75% of crystal having the 55/45 ratio[18]. Similar composition modulations have been reported in polycrystalline samples.[19,6] We expect that most of the scattering stems from the 55/45 composition, however, both compositions are superconducting with the same Tc[18], so the same spectra are expected from both of them. The bulk properties give no indication of more than one superconducting state. So far good superconducting crystals of one composition have not been obtained. The crystal used in the neutron experiment weighed about 11 grams.

Fig.1d shows a scan at the (0.5, 0.5, 0) reciprocal lattice position at 2K and 15K. The measurements were made at the HB-3 triple axis spectrometer at the High Flux Isotope Reactor using collimations of 48-40-80-120 minutes with an initial energy of 14.7 meV using pyrolitic graphite monochromator and analyzer crystals A pyrolitic graphite fillter was placed after the sample to eliminate higher order contamination. The intensity in the neighborhood of 7 meV clearly increases over that at 15K. This is the signature of the resonance excitation that was originally discovered in the cuprate superconductors[14,15,16].Measurements made on polycrystalline $Ba_{0.6}K_{0.4}Fe_2As_2$ showed a spin excitation which appears at the onset of superconductivity[20], and the resonance excitation has been observed in BaFe$_{1.84}$Co$_{0.16}$As$_2$[21] and BaFe$_{1.9}$Ni$_{0.1}$As$^{22}$. Fig.1e shows the temperature dependence of the 7meV scattering. The scattering gets smaller as the temperature gets larger until about 15K or very near T$_c$ as expected for the resonance excitation. Fig.1.f shows a scans at an equivalent crystal wave vectors, but at different momentum transfers. The scattering at (0.5, 1.5, 0) and (1. 5,0.5, 0) is smaller by an amount consistent with the Fe magnetic form factor, but shows the same overall behavior.

Figure 2 shows contour plots of the scattering at three temperatures. The map was obtained by assembling a series of constant energy scans. For the 1.6 K data the scattering is found at the (1/2,1/2, 0) lattice position with the strongest scattering found at about 7 meV which is the position of the resonance. No scattering is observed below about 4 meV showing a gap in the scattering. The scattering near (0.5, 0.5, 0) remains at 20K but is weaker near the resonance position. The scattering now extends down to low energies showing the disappearance of the gap. At 100K the scattering is still visible but spreads out in momentum and is more intense at low energies. No scattering is found at the (0.5, 0, 0) position even though the FeTe parent material shows a signal at that position[23,24]. The Fermi surface for FeTe is very different than that of the superconductor with no appreciable nesting at (Π,Π), so that no scattering is expected there[9]. The Fe(2) sites are occupied to some degree in FeTe but not in the superconductor which may explain the difference in the position of the magnetic scattering.

Fig. 3a. shows that the scattering at 1.6K as a function of (H, -H, 0) averaged over energies from 9 to 13 meV. The scattering appears energy independent in this region which is above the resonance. The scan along (H, -H, 0) makes the trajectory nearly independent of the total momentum Q which is distance from the center of reciprocal space. This gives less importance to Q dependent effects like the form factor and phonon excitations. It does give a larger spectrometer resolution since the scan is largely along the mosaic spread of the sample. However, the mosaic spread is quite good for the basal plane reflections being, on the order of 2 degrees and very symmetric. The c-axis is more disordered in direction as is often the case, so that we will consider this direction when better crystals are available. Scans along the c-axis are not needed to address the central points under consideration. We have checked our results using both the (H, H, 0) basal plane scan and scans with the c axis in the scattering plane. These all gave consistent results within the error of the measurement. Another point that must be considered in dealing with peak widths is that there is a small amount of powder in the sample that did not grow into the crystal. In this case the signal would be centered at the appropriate Q value for the powder and could be wider or narrower then the crystalline peak. The powder scattering is small and can be isolated by performing scans along different paths in the scattering plane. Since Q varies only a small amount for the (H, -H, 0) scan the powder contribution is nearly constant and does not affect the peak shape in any substantial manner. However, care must be taken to understand the role of the powder

scattering since it can affect different scans in different ways.

The curve in Fig.3a is a Gaussian fitted to the red points on the sides of the peak to emphasize the flat top nature of the measured distribution at the top of the peak in black. This flat top is expected in the band picture from excitations between the part of the hole and electron surfaces that are cylindrical and thus give exact $(\Pi,\Pi)$ transitions[9,10,11,12]. The red points give results from transitions between hole and electron sheets where the sheets differ in size or shape so the transitions are not quite $(\Pi,\Pi)$. Data taken at 15K are essentially the same so are not shown here.
In the band picture this width stems from the imperfect nesting between the hole and electron sheets of the band structure. It could stem from quite different shapes of the Fermi surface such as those discovered in Ref.13. Fig.3b shows the same scan taken at an energy transfer of 7 meV which is the energy of the resonance. The signal near the center of the 1.6K distribution increases below Tc and stems from transitions with a wave vector exactly at $(\Pi,\Pi)$ which gives the resonance. However, there is another set of excitations that are not visibly changed below Tc. These come from transitions away from $(\Pi,\Pi)$ so they do not show up in the resonance. These transitions are found above Tc so are available to provide the pairing and have a large integral weight compared to the resonance. The excitations at $(\Pi,\Pi)$ stem from a region of instability that can favor a spin or charge density wave as can be seen in the recent photoemission work[13]. Here the resonance is found at that position and prior work has shown that the resonance itself is associated with a small magnetic energy compared to that needed for pairing[26]. The large density of excitations not in the resonance can provide the necessary energy. The peak width is limited by the spectrometer resolution for the scans in Fig. 3. Resolution calculations confirmed by scans in different directions suggest that the actual width is 70±10% of the width shown.

This work shows how superconductivity occurs in an itinerant electron picture. The Fermi surface consists of hole and electron surfaces that resemble cylinders and are arranged so that $(\Pi,\Pi)$ transitions take place between the cylinders giving the flat topped spectra seen in Fig. 3a. Such a Fermi surface would be unstable against spin or charge formation. However, there are many more near $(\Pi,\Pi)$ transitions made possible by the fact that the hole and electron surfaces are not perfect identical cylinders and these favor pairing[10]. These produce the wide part of the scattering shown by the region inside the fitted curves but outside the area where perfect $(\Pi,\Pi)$ transitions are possible. At the typical energy[27] of about 6Tc the $(\Pi,\Pi)$ transitions produce the resonance for temperatures below Tc.

The portions of this work conducted at Oak Ridge National Laboratory were supported by the Scientific User Facilities Division and by the Division of Materials Science and Engineering, Office of Basic Energy Sciences, DOE.

**Figure Captions.**

Figure 1 Properties and Neutron Scans for Fe(TeSe). a, Crystal structure of the Fe(TeSe) material showing the planer nature of the material and the two Fe positions. b, . Magnetic susceptibility of a FeTe$_{0.5}$Se$_{0.5}$ crystal measured with H = 20 Oe using zfc and fc protocols. The diamagnetic susceptibility for the zfc data corresponds to complete diamagnetic screening. The resistivity data from the same sample are also shown. c, Estimation of the electronic contribution to the heat capacity of a FeTe$_{0.5}$Se$_{0.5}$ crystal near and below T$_c$ ≈ 14 K. This crystal clearly shows bulk superconductivity. The lattice contribution to the heat capacity of FeTe$_{0.5}$Se$_{0.5}$ is estimated by measuring the heat capacity of a FeTe$_{0.67}$Se$_{0.33}$ crystal that shows no evidence of bulk superconductivity. The low temperature heat capacity data of the FeTe$_{0.67}$Se$_{0.33}$ crystal is accurately described by γT + BT$^3$+CT$^5$, with γ = 39 mj/mole-K$^2$ and Θ$_D$ = 174 K. The heat capacity data from the FeTe$_{0.5}$Se$_{0.5}$ crystal are adjusted by a few percent to match the FeTe$_{0.67}$Se$_{0.33}$ data at T= 20 K and the estimated lattice contribution (BT$^3$ + CT$^5$) subtracted. d, data taken at 1.6 and 20K. The low temperature curve rises about the high temperature curve the most at about 7meV giving the resonance energy. e. Temperature dependence of the scattering at the resonance position. f. Comparison of data taken at a larger Q position to show the reduction in signal caused by the magnetic form factor. This proves the scattering is magnetic.

Figure 2 Contour plots showing the behavior of the magnetic scattering at three temperatures. The 1.6K data shows the strong scattering near 7 mev stemming from the resonance. A gap in the scattering appears below about 4 meV. The gap vanishes at 20K and the scattering in the resonance region is reduced. However, a lot of intensity remains. At 100K the spectra spreads out intensity in both energy and momentum. Some intensity appears near (1,1,0), but the origin of the scattering is not known. It may stem from phonons.

Figure 3 Scans through (0.5, -0.5, 0) for energies above the resonance and at the resonance. Fig.3a data at 9 11 and 13 meV averaged and plotted for 1.6K. The scan has a flat top so the data were fitted using the red points only. b Scans taken at 7 meV for temperatures above and below Tc.

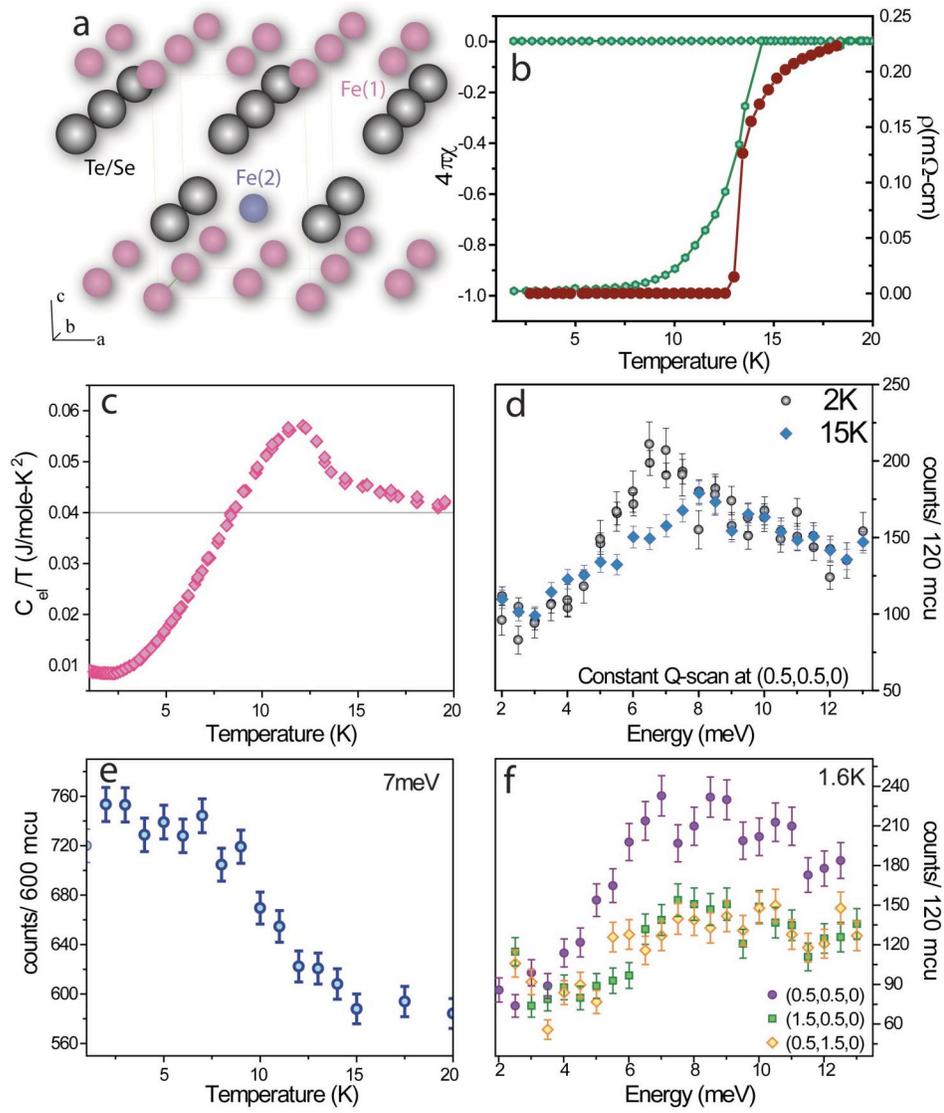

**Fig. 1**

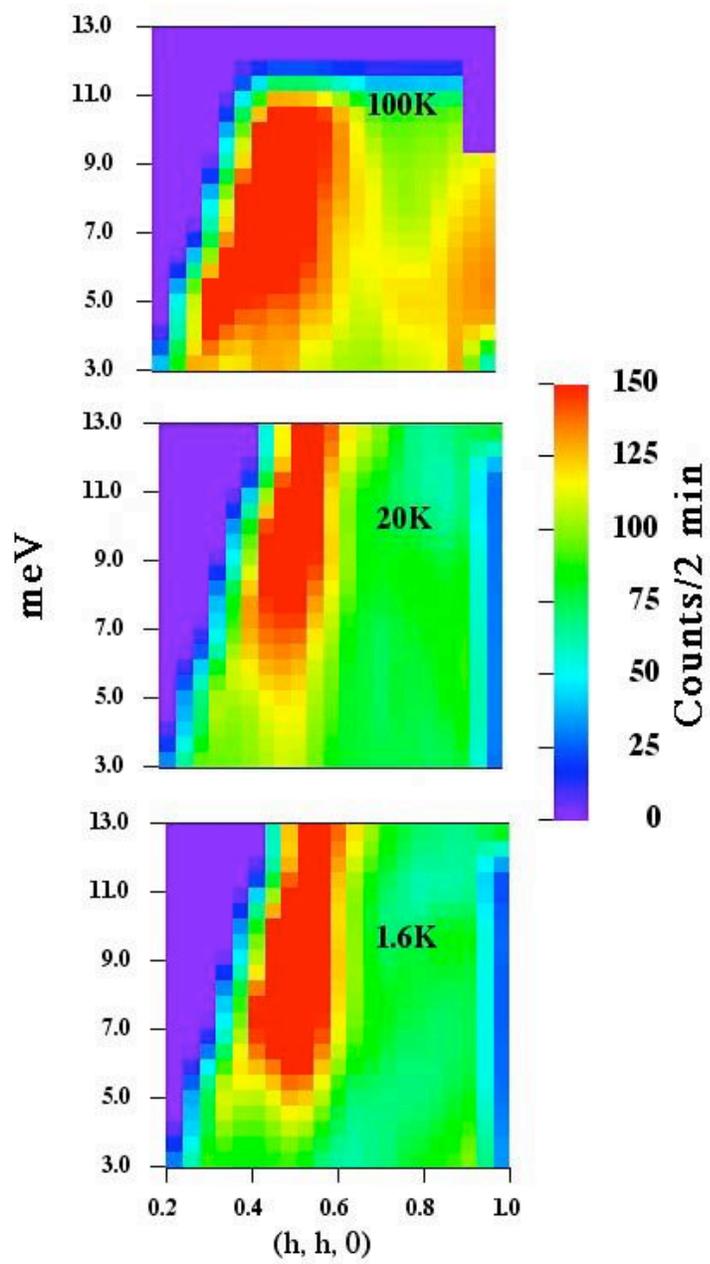

**Fig. 2**

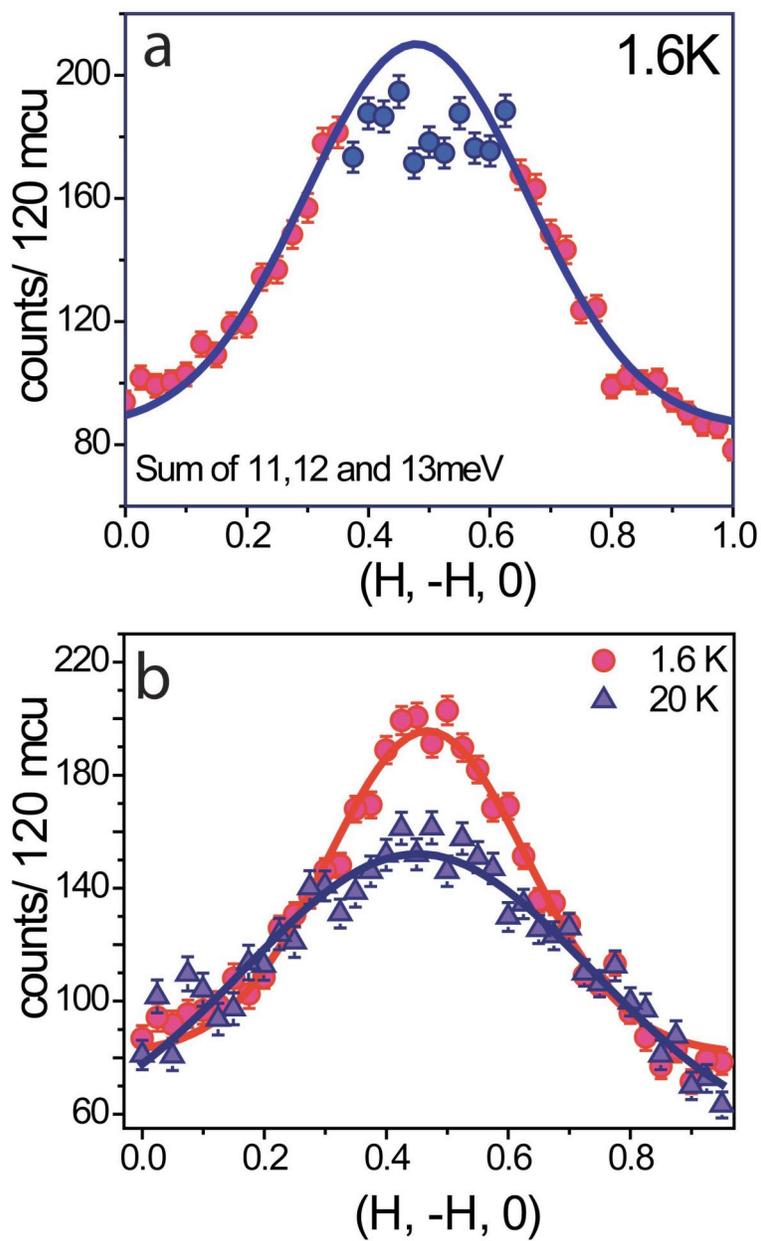

**Fig. 3**